\documentstyle[prl,pra,multicol,aps,epsf]{revtex}
\draft
\begin{document}
\draft
\pagenumbering{roma}
\title{Semiconductor-cavity  QED in high-Q regimes: Detuning effect}

\author{Yu-xi Liu,$^{(a)}$ N. Imoto,$^{(a,b,c)}$  and \c{S}. K. \"Ozdemir$^{(a,b)}$}
\address{$(a)$The Graduate University for Advanced Studies (SOKEN),
  Hayama, Kanagawa, 240-0193, Japan \\
  $(b)$ CREST Research Team for Interacting Carrier Electronics\\
  $(c)$ NTT Basic Research Laboratories, 3-1 Morinosato Wakamiya, Atsugi, Kanagawa,
  243-0198, Japan}
\author{Guang-ri Jin, C. P. Sun}
\address{Institute of Theoretical Physics, The Chinese Academy of
Sciences, P. O. Box 2735, Beijing 100080, China}

\maketitle \vspace{1mm} \widetext
\begin{abstract}
The non-resonant interaction between the high-density excitons in
a quantum well and a single mode cavity field is investigated. An
analytical expression for the physical  spectrum of the excitons
is obtained. The spectral properties of the excitons, which are
initially prepared in the number states or the superposed states
of the two different number states by the resonant femtosecond
pulse pumping experiment, are studied. Numerical study of the
physical spectrum is carried out and a discussion of the detuning
effect is presented.
\end{abstract}
\widetext
\begin{multicols}{2}
\pagenumbering{arabic}
\section{introduction}

The optical properties of the semiconductor structures have been
the subject of intensive experimental and theoretical
investigation during recent years. Technical progress in the
semiconductor crystal growth  has made it possible to control
semiconductor structures in all three spatial dimensions.
Experimentally,  physicists can fabricate the quantum wells, in
which a thin semiconductor film is sandwiched between different
materials via  heterojunctions so that the motion of carriers is
confined in the two dimensional thin-film plane. This confinement
of carriers to two dimensions gives rise to new quantum effect not
observed in bulk materials, for example, an electric field-induced
energy shift of the resonance which is called the quantum confined
Stark effect~\cite{miller}. Multi-dimensional quantum confined
structure, such as quantum wires and quantum dots, are expected to
further improve the quantum  effect of the optical devices.
Experiments have shown that the confinement of carriers in these
low dimensional semiconductor structures (LDSS's)  can result in
novel optical-electronic effects which may lead to fabrication of
new optical components. Some new fascinating achievements in the
areas,  such as semiconductor microcavity (SMC) quantum
electrodynamics (QED), quantum dot microlaser and turnstile
device~\cite{A}, quantum computer with quantum dot~\cite{B,aa},
and semiconductor random laser~\cite{C} etc., encourage the study
of optical-electronic properties of these micro-structures.

The present trend toward smaller-scale nanostructures and the
continuous development of new and improved materials have started
a steady progress towards fabrication of more ideal optical
microcavities. If an LDSS is placed in an SMC, the optical mode
structure of the SMC will change  around the LDSS.  Using this
effect many interesting phenomena, such as tailoring the
spontaneous radiation pattern and rate~\cite{1,2,3}, the coupled
exciton-photon mode splitting  in a semiconductor quantum
microcavity~\cite{4}, have been demonstrated. Therefore the
investigation of  optical properties of an LDSS placed into a
semiconductor cavity is very important and necessary for
theoretical and  experimental physicists.

It is known that the interaction between the light and these
micro-systems  occurs via exciton~\cite{D,E} which is an
electron-hole pair bound by the Coulomb attraction. The radiation
of exciton exhibits the superradiant character. The initial
theoretical~\cite{M} and experimental~\cite{N} studies were
focused on the superradiance of the Wannier excitons in
semiconductor micro-crystallites. Then, the superradiance  of the
Frenkel excitons was observed in J aggregates at  low
temperature~\cite{O,P}.  In 1995, the superradiance of high
density Frenkel excitons in  a R-Phycoerythrin(R-PE) single
crystal was observed at room temperature for the first
time~\cite{R}. We have studied the spontaneous radiation of the
Frenkel excitons in a crystal slab under the condition of low
excitation with an exactly solvable model and shown its
superradiance nature~\cite{Q}. We have also discussed the quantum
statistical properties of the output field and the semiconductor
QED for the high density excitons in a semiconductor
microcavity~\cite{5,Liu}. In the former works, our main interest
was  the resonant interaction between the excitons and the cavity
field. But we know that the detuning between the cavity field and
exciton always affects the radiation properties of the exciton in
the quantum well~\cite{coch}. So in this paper we will discuss a
general model of the interaction between the high density excitons
in a quantum well and a single-mode cavity field.

 In section II, we give  a general theoretical model of the
non-resonant interaction between a single-mode cavity field and the exctions. By virtue of the Schwinger's
representation of the angular momentum using two boson modes, we  approximately obtain the analytical
solution of the system.  In section  III, the stationary physical  spectrum of  the excitons,  which are
initially in the number state or the superposed state of two different number states is presented. A
comparison of these results with those of the resonant  cases will be also given.  Finally, a brief summary
and conclusion are presented in section IV.

\section{theoretical model and analytical solution}
In this section, we present a theoretical model to study  the
interaction  between  a quantum well and a single-mode cavity
field. We assume that the cavity and the quantum well are ideal,
and they are in an extremely low temperature circumstance. The
quantum well interacts with cavity field via exciton, which is an
electron-hole pair bound by the Coulomb attraction. At extremely
low temperatures, the thermal momentum of the excitons is so small
that the thermalized excitons can be neglected. Then there are
only excitons with zero in-plane momentum. It is well known  that
when the density of the excitons becomes high, the ideal bosonic
model of the excitons is no longer adequate (In the case of a GaAs
quantum well, the ideal bosonic model becomes inadequate when the
density of the excitons exceeds $1.3 \times 10^{9}$
cm$^-2$~\cite{66} ) for a theoretical study. However, we can
describe exciton operators as hypothetical bosonic operators. In
order to deal with the deviation of the exciton operators from the
ideal bosonic operators, we introduce an effective non-linear
interaction between these hypothetical ideal bosons to correct the
high excitons density effect~\cite{6,hanamura}. These
considerations lead to the following Hamiltonian after the
rotating-wave approximation is made~\cite{Liu,6}:
\begin{equation}
H=H_{0}+H^{\prime}
\label{eq:H}
\end{equation}
with
\begin{mathletters}
\begin{eqnarray}
H_{0}&=&\hbar\omega_{1}a^{\dagger}a+\hbar\omega_{2}b^{\dagger}b+
\hbar g(a^{\dagger}b+b^{\dagger}a),
\label{eq:h0}  \\
H^{\prime}&=&\hbar A b^{\dagger}b^{\dagger}bb-\hbar \nu
(a^{\dagger}b^{\dagger}bb+b^{\dagger}b^{\dagger}ab),
\end{eqnarray}
\end{mathletters}
where $b^{\dagger}(b)$ are creation (annihilation) operators of the excitons with frequency $\omega_{2}$, and
$a^{\dagger}(a)$ are the creation (annihilation) operators of the cavity field with frequency $\omega_{1}$.
We assume that both of them  obey the bosonic commutation relation $[b, b^{\dagger}]=[a, a^{\dagger}]=1$. $g$
stands for the interaction strength between the cavity field and the excitons. $A$ represents an effective
interaction constant between the excitons. $A$ is assumed as a positive real number, which means that the
bi-excitons are not stable in this system of the quantum well. So, it is reasonable to consider that only
excitons are present in the quantum well.  $\nu$ is a positive real number which describes the phase space
filling factor. The ratio of the phase space filling factor to the interaction constant of the excitons is
about $\nu/A=0.3$~\cite{jii}. For the sake of simplicity, we assume that the parameters $A$ and $\nu$ are of
the same order, or $A^2<\nu$. In that case, it is convenient to give the solution of eq.(\ref{eq:h0}) using
the Schwinger's representation of the angular momentum for two-mode bosonic operators~\cite{lc}. Angular
momentum operators can be constructed as
\begin{mathletters}
\begin{eqnarray}
J_{x}&=&\frac{1}{2}(a^{\dagger}b+b^{\dagger}a),
J_{y}=\frac{1}{2i}(a^{\dagger}b-b^{\dagger}a), \nonumber \\
J_{z}&=&\frac{1}{2}(a^{+}a-b^{+}b),
\end{eqnarray}
and total number operator as
\begin{equation}
\hat{N}=a^{+}a+b^{+}b,
\end{equation}
\end{mathletters}
using the ladder operators $a^{\dagger}, a$ of the cavity field and the exciton operators $b^{\dagger}, b$.
It is obvious that $\hat{N}$ is the constant of motion with respect to the Hamiltonian (\ref{eq:H}). Then the
total angular momentum operator can be expressed as
\begin{equation}
J^{2}=J^{2}_{x}+J^{2}_{y}+J^{2}_{z}=\frac{\hat{N}}{2}(\frac{\hat{N}}{2}+1).
\end{equation}
For any fixed total particle number $\cal{N}$, the common eigen-states
of $J^{2}$ and $J_{z}$ are
\begin{equation}
|jm\rangle
 =\frac{(a^{\dagger})^{j+m}(b^{\dagger})^{j-m}}{\sqrt{(j+m)!(j-m)!}}
|0\rangle,
\end{equation}
with the eigenvalues $j=\frac{\cal{N}}{2}$, and $ m=-\frac{\cal{N}}{2},\ldots, \frac{\cal{N}}{2}$, where
$|jm\rangle$ are the Fock states with $j+m$ photons in the cavity and $j-m$ excitons in the quantum well,
respectively. Although $j\pm m$ must be integers, $j$ and $m$ can  both be integers or both be half-odd
integers. If we define
\begin{equation}
\Omega=\frac{1}{2}(\omega_{1}+\omega_{2}),
\Delta=\frac{1}{2}(\omega_{1}-\omega_{2}),
\end{equation}
then eq.(\ref{eq:h0}) can be simplified  into
\begin{eqnarray}
H_{0}&=&\hbar\Omega\hat{N}+
\hbar G(sin \theta J_{x}+cos\theta J_{z}) \nonumber \\
&=&\hbar \hat{N}+\hbar G e^{-i\theta J_{y}} J_{z}e^{i\theta J_{y}}
\label{eq:sh}
\end{eqnarray}
in terms of an $SO(3)$ rotation $e^{i\theta J_{y}}$ of
$\hbar\Omega\hat{N}+\hbar G(sin \theta J_{x}+cos\theta J_{y})$
with $G=\sqrt{\Delta^{2}+4g^{2}}$ and
$tg\theta=\frac{2g}{\Delta}$. The eigenfunctions and eigenvalues
of Hamiltonian (\ref{eq:sh}) are, respectively,
\begin{equation}
|\psi^{(0)}_{jm}\rangle=e^{-i\theta J_{y}}|jm\rangle,
E^{(0)}_{jm}=\hbar({\cal N}\Omega+mG).
\label{eq:pp}
\end{equation}
It is clear that eigenfunction $|\psi^{(0)}_{jm}\rangle$ represents a dressed exciton state or a polariton
state. Based on eq.(\ref{eq:pp}), using perturbation theory and keeping $A$ and $\nu$ parameters up to their
first-order, we obtain the eigenvalues of the Hamiltonian (\ref{eq:H}) as:
\begin{mathletters}
\begin{equation}
E_{jm}=E^{(0)}_{jm}+\langle jm|e^{i\theta J_{y}} H^{\prime}
e^{-i\theta J_{y}}|jm\rangle,
\end{equation}
and
\begin{equation}
|\psi_{jk}\rangle=|\psi^{(0)}_{jk}\rangle+\sum_{n\ne k}
\frac{\langle\psi^{(0)}_{jn}|H^{\prime}
|\psi^{(0)}_{jk}\rangle}{E^{(0)}_{jk}-E^{(0)}_{jn}}
|\psi^{(0)}_{jn}\rangle.
\label{eq:ee}
\end{equation}
\end{mathletters}

In order to obtain $E_{jm}$ and $|\psi_{jk}\rangle$, we first calculate the matrix elements of the
perturbation term as follows
\end{multicols}
\widetext
\begin{eqnarray}
&&\langle\psi^{(0)}_{jn}|H^{\prime}|\psi^{(0)}_{jm}\rangle=
\hbar[A(\cos\frac{\theta}{2})^{4}+\nu \sin\theta (\cos\frac{\theta}{2})^2]
(j-m-1)(j-m)\delta_{n,m}+\hbar[A(\sin\theta)^{2}-\nu \sin 2\theta]
(j^{2}-m^{2})\delta_{n,m} \nonumber \\
&&+ \hbar[A(\cos\frac{\theta}{2})^{2}\sin \theta
-\nu \cos\theta (\cos\frac{\theta}{2})^2+\frac{\nu}{2}(\sin\theta)^2]
(j-m)\sqrt{(j+m)(j-m+1)}\delta_{n,m-1} \nonumber\\
&&+\hbar[\frac{A}{4}(\sin\theta)^{2}
-\frac{\nu}{4}\sin 2\theta]
\sqrt{(j+m)(j+m-1)}\sqrt{(j-m+1)(j-m+2)}\delta_{n,m-2} \nonumber \\
&&+\hbar[A\sin\theta (\cos\frac{\theta}{2})^{2}-\nu \cos\theta
(\cos\frac{\theta}{2})^2+\frac{\nu}{2} (\sin \theta)^2]
\sqrt{(j-m)(j+m+1)}(j-m-1)\delta_{n,m+1} \nonumber \\
&&+\hbar[A\sin\theta(\sin\frac{\theta}{2})^{2}-\nu \cos\theta
(\sin\frac{\theta}{2})^2-\frac{\nu}{2}(\sin\theta)^2]
\sqrt{(j+m)(j-m+1)}(j+m-1)\delta_{n,m-1} \nonumber \\
&&+\hbar[A\sin\theta(\sin\frac{\theta}{2})^{2}-\nu \cos\theta
(\sin\frac{\theta}{2})^2-\frac{\nu}{2}(\sin\theta)^2]
\sqrt{(j-m)(j+m+1)}(j+m)\delta_{n,m+1}\nonumber \\
&&+\hbar[\frac{A}{4}(\sin\theta)^{2}-\frac{\nu}{4}\sin 2\theta]
\sqrt{(j+m+1)(j+m+2)}\sqrt{(j-m)(j-m-1)}\delta_{n,m+2}\nonumber \\
&&+\hbar[A(\sin\frac{\theta}{2})^{4}-\nu
\sin\theta(\sin\frac{\theta}{2})^2](j+m)(j+m-1)\delta_{n,m},
\label{eq:pp1}
\end{eqnarray}
\widetext
\begin{multicols}{2}\noindent
and then write the eigenvalues of the Hamiltonian (\ref{eq:H})as:
\begin{eqnarray}
&&E_{jm}=\hbar\Omega {\cal N}+\hbar G m+
\hbar[A(\sin\theta)^{2}-\nu \sin 2\theta ]
(j^{2}-m^{2}) \nonumber \\
&&+\hbar[A(\cos\frac{\theta}{2})^{4}+\nu \sin\theta (\cos\frac{\theta}{2})^2]
(j-m-1)(j-m) \nonumber \\
&&+\hbar[A(\sin\frac{\theta}{2})^{4}-\nu
\sin\theta(\sin\frac{\theta}{2})^2](j+m)(j+m-1).
\end{eqnarray}
All the eigenfunctions of the Hamiltonian (\ref{eq:H}) can be easily obtained using eq.(\ref{eq:ee}) and
eqs.(\ref{eq:pp},\ref{eq:pp1}). Then  the time evolution operators of the system  can be written as
\begin{equation}
U(t)=e^{-\frac{i}{\hbar}H t}=\sum_{j=0}^{\infty}\sum_{m=-j}^{j}
e^{-it\frac{E_{jm}}{\hbar}}|\psi_{jm}\rangle\langle \psi_{jm}|.
\label{eq:ef}
\end{equation}
Consequently, for any initial state $|\psi(0)\rangle$, the time-dependent wave function can be obtained as
$|\psi (t)\rangle=U(t)|\psi (0)\rangle$.

\section{radiation spectrum of exciton}

Under the condition  of ideal  cavity, ideal quantum well, and
extremely low temperature, both the excitons and the cavity field
have zero linewidth.  Assumption of ideal quantum well is
important to eliminate  the linewidth of excitons which may be
caused by the fluctuations of the quantum well. So  the only
broadening mechanism comes from the detecting spectrometer for
which the physical spectrum can be defined as ~\cite{jh}
\begin{equation}
S(\omega)=2\gamma\int_{0}^{t}{\rm d}t_{1}\int_{0}^{t}{\rm d}t_{2}
e^{-(\gamma-i\omega)(t-t_{2})}e^{-(\gamma+i\omega)(t-t_{1})}G(t_{1},t_{2})
\end{equation}
where $\gamma$ is the half-bandwidth of the spectrometer, and $t$
is the time length of the excitation in the cavity.
$G(t_{1},t_{2})$  represents the dipole correlation function of
the excitons and is defined as
\begin{equation}
G(t_{1},t_{2})=\langle\psi(0)|B^{\dagger}(t_{2})B(t_{1})|\psi(0)\rangle
\end{equation}
with the initial state $|\psi(0)\rangle$ of the system,  and
$B^{\dagger}=b^{\dagger}-\nu b^{\dagger}b^{\dagger}b$ $(B=b-\nu
b^{\dagger}bb)$ where the second term comes from the correction of
the phase space filling effect~\cite{tohya}. Taking into account
the fact that $b(t)=U^{\dagger}(t)bU(t)$, we can obtain the
correlation $G(t_{1},t_{2})$ as
\begin{eqnarray}
G(t_{1},t_{2})&=&\langle\psi(0)|U^{\dagger}(t_{2})B^{\dagger}U(t_{2})
U^{\dagger}(t_{1})BU(t_{1})  |\psi(0) \rangle \nonumber \\
&=&\sum_{j,k,l,m,n}\langle\psi_{jl}|B^{\dagger}|\psi_{km} \rangle
\langle\psi_{km}|B|\psi_{jn} \rangle \nonumber \\
&\times& \langle\psi(0)|\psi_{jl}\rangle
\langle\psi_{jn}|\psi(0) \rangle e^{i\omega_{jl,km}t_{2}}
 e^{-i\omega_{jn,km}t_{1}}
\end{eqnarray}
with $\omega_{jl,km}=(E_{jl}-E_{km})/\hbar$ and $\omega_{jn,km}=(E_{jn}-E_{km})/\hbar$. It is evident that
$j$ is determined only by the initial state $|\psi(0)\rangle$. Then the stationary physical spectrum can be
written  as
\begin{eqnarray}
S(\omega)&=&\sum_{j,l,k,m}\frac{2\gamma}{\gamma^{2}+
(\omega-\omega_{jl,km})^{2}} \nonumber \\
&\times&|\langle\psi(0)|\psi_{jl}|^{2}|\langle\psi_{jl}|B^{\dagger}
|\psi_{km}\rangle|^{2}. \label{eq:s}
\end{eqnarray}
In order to fulfill the condition $\langle m^{\prime}j^{\prime}|B^{\dagger}|jm\rangle\ne 0$, both
 $j^{\prime}=j+\frac{1}{2}$ and $m^{\prime}=m-\frac{1}{2}$
 must be satistied. According to the above selection
rule, eq.(\ref{eq:s}) can be rewritten as
\begin{eqnarray}
S(\omega)&=&\sum_{j,l,m}\frac{2\gamma}{\gamma^{2}+
(\omega-\omega_{jl,j-\frac{1}{2}m})^{2}} \nonumber \\
&\times&|\langle\psi(0)|\psi_{jl}\rangle|^{2}|\langle\psi_{jl}|B^{\dagger}
|\psi_{j-\frac{1}{2}m}\rangle|^{2}.
\label{eq:f}
\end{eqnarray}
The eigenvalues determine the position of the spectral component and
$|\langle\psi(0)|\psi_{jl}|^{2}|\langle\psi_{jl}|B^{\dagger} |\psi_{j-\frac{1}{2}m}\rangle|^{2}$ determine
the intensity of the spectral  lines. The above spectral formula is similar to that of reference~\cite{Liu}.

In the following, we will consider that the system is initially
prepared in the bare exciton state using the resonant femtosecond
pulse pumping method~\cite{Cao}. For simplicity, we assume that
$\hbar=1$, so  that all frequency quantities  have a unit of
energy. Moreover, we impose that initially there are no photons in
the cavity. Then for a system, which is initially prepared in the
single exciton state, the physical spectrum from ${\cal N}=1$ to
${\cal N}=0$ can be simplified into
\begin{eqnarray}
S(\omega)&=&\frac{2\gamma (sin\frac{\theta}{2})^{4}}
{\gamma^{2}+(\omega-\Omega-\frac{G}{2})^2  } \nonumber\\
&+&\frac{2\gamma (cos\frac{\theta}{2})^{4}}
{\gamma^{2}+(\omega-\Omega+\frac{G}{2})^2} \label{eq:sp}
\end{eqnarray}
which has double peaks located at $\Omega+\frac{G}{2}$  and $\Omega-\frac{G}{2}$.  It is clear that the phase
space filling factor and the interaction between the excitons, which are the results of the multi-exciton
presence, don't affect the physical spectrum of this system.

%\vspace{0.5cm}
\begin{figure}
\epsfxsize=7.0cm \centerline{\epsffile{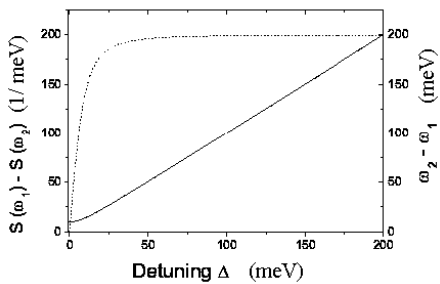}}

\vspace{1cm} \caption[]{$1000$ times of the difference between the
heights\\ of two peaks (dot line),  and the difference between
positions\\ (solid line) of two peaks are plotted as function of
the\\ detuning $\Delta$ for $\gamma=0.01$ meV, $g=6$ meV.
$\omega_{1}=\omega-\Omega=-\frac{G}{2}$ \\and
$\omega_{2}=\omega-\Omega=\frac{G}{2}$.}
\end{figure}

Figure $1$ depicts the effect of detuning $\Delta$ on the relative
heights and the positions of the two peaks in the spectrum of the
system described by eq. (\ref{eq:sp}).  The difference between the
heights of the two peaks in the physical spectrum is zero when
$\Delta=0$ (resonance case). This profile line is observed for
two-level atomic systems, too. In the non-resonant case, peaks
have different heights and the frequency difference between them
become larger due to detuning effect. With the increase of
$\Delta$, the height of the peak locating at $\Omega+\frac{G}{2}$
reduces gradually, whereas the height of the peak at
$\Omega-\frac{G}{2}$ increases. From eq. (\ref{eq:sp}) and the
condition $tg\theta=2g/\Delta$, it can be derived that if $\Delta$
and the coupling constant $g$ satisfy the condition $\Delta\gg
2g$, then $\theta\approx 0$, which means the physical spectrum has
approximately one peak. It is seen in Fig. $1$ that when
$\Delta\geq10g$, the difference between the heights of the peaks
becomes constant. A further analysis will reveal that at those
values of $\Delta$, the height of the peak located at
$\Omega+\frac{G}{2}$ becomes very close to zero, therefore the
spectrum can be interpreted as having a single peak.

With the increase of the pumping power, the emission spectrum of the system becomes complex. The simplest
non-trivial example, for which the non-linear terms play a role in the physical spectrum, is obtained when
the quantum well has initially two excitons and there is no photon in the cavity. In that case,  the physical
spectrum from ${\cal N}=2$ to ${\cal N}=1$ becomes
\begin{equation}
S(\omega)=\sum_{l,m}\frac{2\gamma}{\gamma^{2}+
(\omega-\omega_{l,m})^{2}}
|\langle\psi(0)|\psi_{1l}\rangle|^{2}|\langle\psi_{1l}|B^{\dagger}
|\psi_{\frac{1}{2}m}\rangle|^{2}
\label{eq:pppp}
\end{equation}
with $\omega_{lm}=\omega_{1l,\frac{1}{2}m}$. We know that there
are three and two dressed exciton states for ${\cal N}=2$ and
${\cal N}=1$, respectively. According to the selection rule, there
should be six peaks in the physical spectrum (\ref{eq:pppp}).
 \begin{figure}
\epsfxsize=8cm \centerline{\epsffile{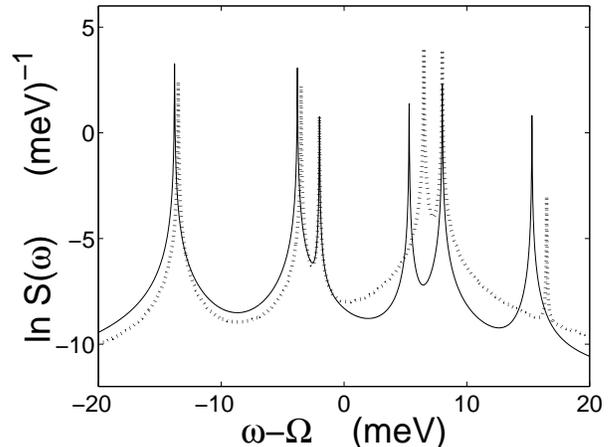}}
\caption[]{$S(\omega)$ are ploted as a function of the frequency
$\omega-\Omega$\\  for $g=5$ meV, $A/g=0.6$, $\gamma=0.01$ meV
when (a)\\ $\nu=0$ (dashed line),  and (b) $\nu/A=0.3$ (solid
line).}
\end{figure}
Figure $2$, which is drawn using  eq.(\ref{eq:pppp}), depicts the
radiation spectrum $S(\omega)$ of the excitons as a function of
the frequency difference $\omega-\Omega$
 under the resonant condition with or without considering the phase
space filling factor. In order to give a clear picture, the natural logarithm of the physical spectrum is
plotted in Fig.$2$ (in Figs.$3$ and $6$, as well). It is understood that the phase filling factor does not
affect the line profile of the spectrum, but changes the heights and the positions of the peaks. So in the
following numerical results, we have neglected the phase space filling effect.

Figure 3 clarifies that the detuning between the cavity field and
the excitons changes not only the positions of the peaks, but also
reduces their heights. The larger the detuning is, the stronger
its effect on the physical spectrum is. As it is seen in Fig.$3$,
$\Delta=2$ meV only slightly changes the position and heights of
the peaks keeping their number constant. However, with increasing
detuning up to $\Delta=200$ meV, the number of the peaks in the
spectrum gradually decreases because (i) the heights of some peaks
become so small when compared to main peak that they can be
neglected, and (ii) some peaks are shifted so close to each other
that they can not be resolved.

\begin{figure}
\epsfxsize=8 cm \centerline{\epsffile{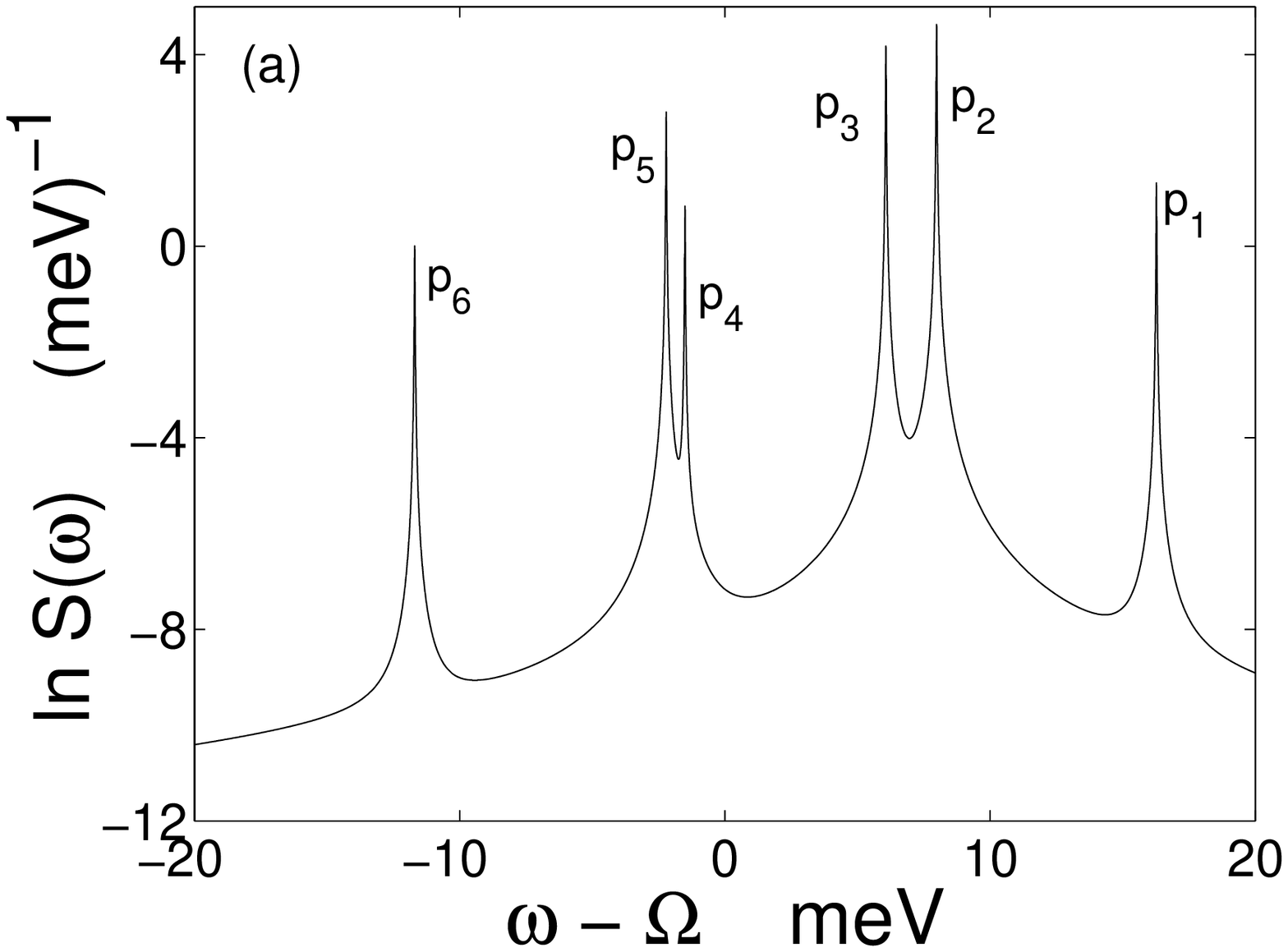}} \epsfxsize=8cm
\centerline{\epsffile{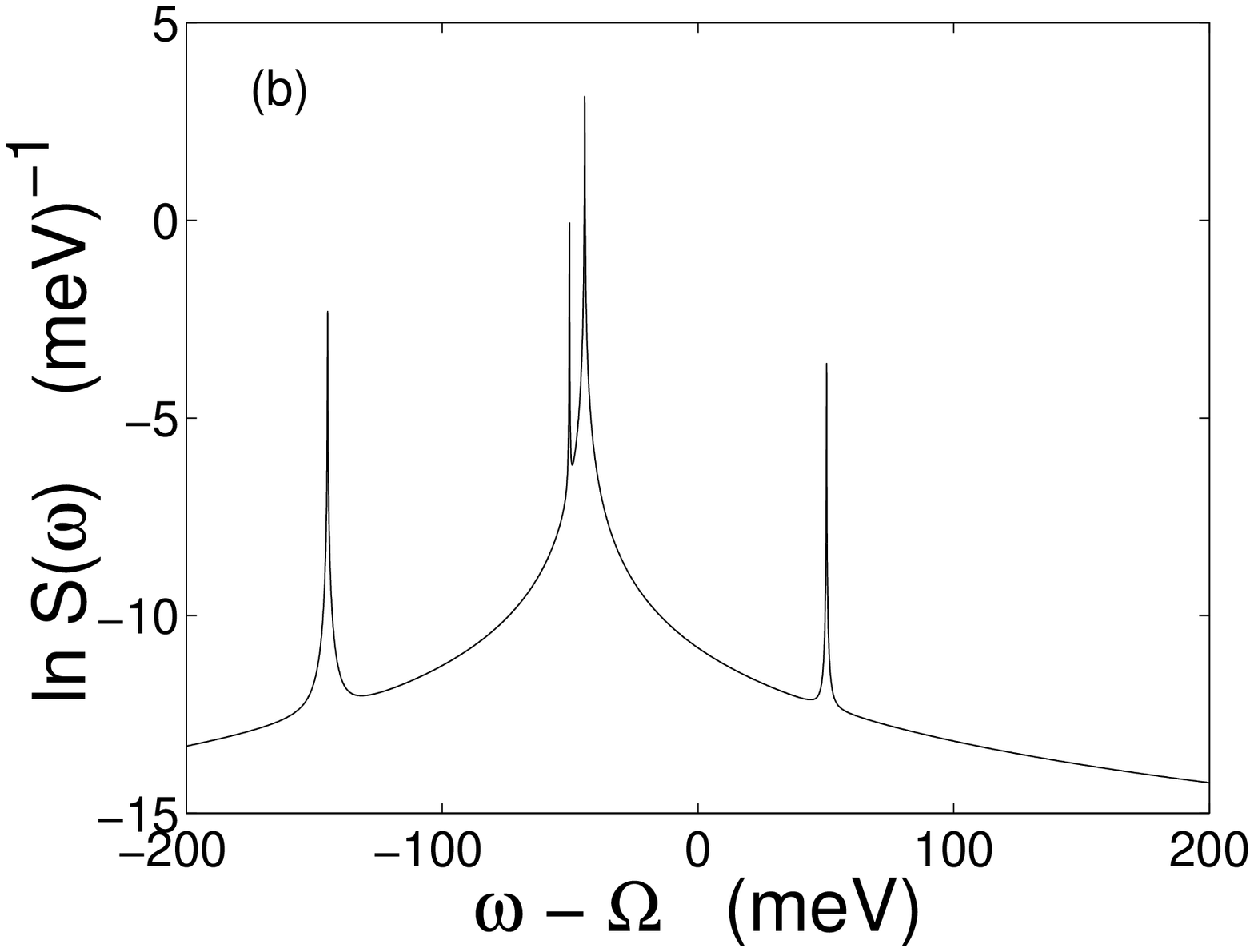}} \epsfxsize=8 cm
\centerline{\epsffile{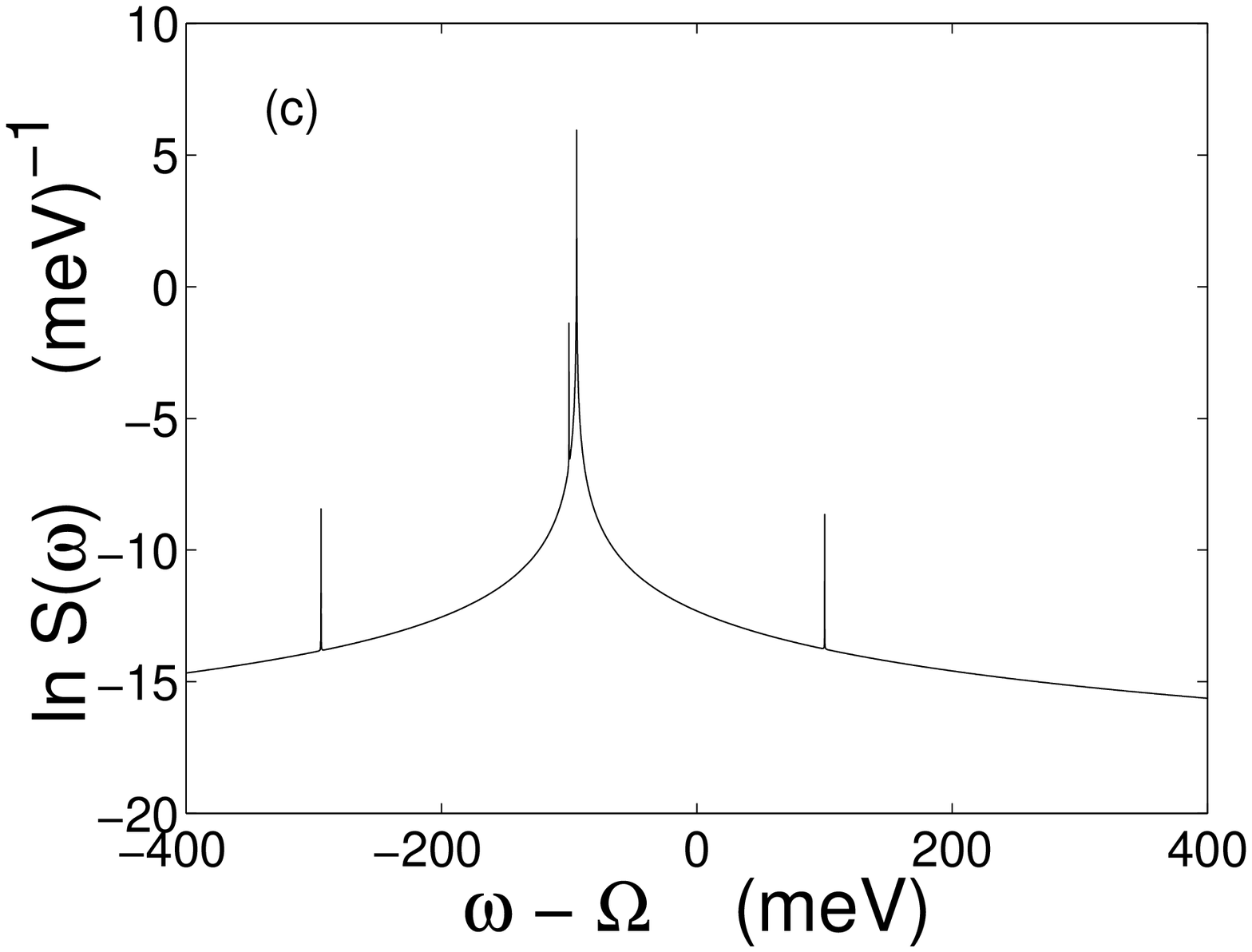}} \caption[]{$S(\omega)$ is plotted
as a function of the frequency \\$\omega-\Omega$  for a set of
parameters $g=5$ meV, $A/g=0.6$, \\  $\gamma=0.01$ meV (a)
$\Delta=2$ meV (b), $\Delta=100$ meV \\(c), $\Delta=200$ meV.}
\end{figure}

The  variation of the positions of the six peaks (labeled as
$p_{1}$, $p_{2}....p_{6}$ in the spectrum of Fig.$3a$) of
eq.(\ref{eq:pppp}) is plotted as a function of detuning $\Delta$
in Figure $4$. It is seen that with the increase in detuning, the
difference between $p_{2}$ and $p_{3}$ becomes almost zero so that
they can not be resolved. Peaks $p_{1}$ and $p_{6}$, respectively,
shifts to more positive and negative sides of the spectrum. The
positions of $p_{4}$ and $p_{5}$ are interchanged with increasing
$\Delta$, and after this interchange their relative positions are
kept the same.

\begin{figure}
\epsfxsize=8cm \centerline{\epsffile{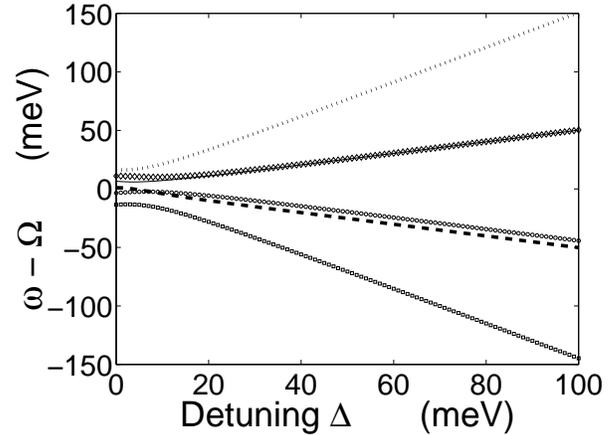}} \caption[]{The
positions of peaks are plotted as function of\\
detuning $\Delta$ for $\gamma=0.01$ meV, $g=5$ meV, $A/g=0.6$.\\
From above line to below line, they correspond respectively \\to
the positions of $p_{1}$ (dot line), $p_{2}$ (diamond  line),\\
$p_{3}$ (solid line), $p_{4}$ (dashing line), $p_{5}$ (circle
line), \\$p_{6}$ (square line).}
\end{figure}

\begin{figure}
\epsfxsize=8 cm \centerline{\epsffile{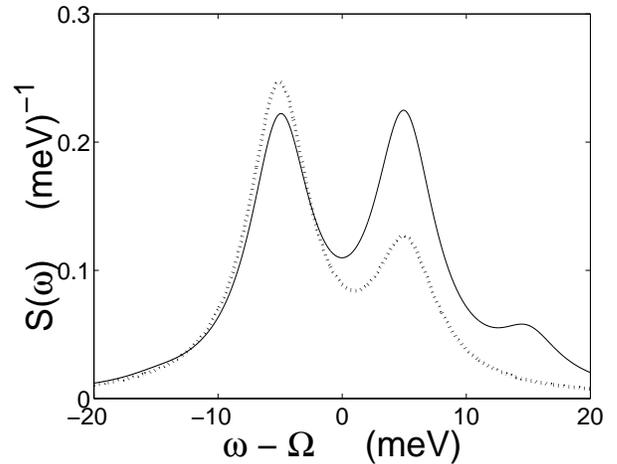}}
\caption[]{$S(\omega)$ is ploted as a function of the frequency
$\omega-\Omega$\\ for $g=5$ meV, $A/g=0.001$, $\gamma=3$ meV  when
(a)\\$\Delta=0$ (solid line), and (b) $\Delta=2$ meV (dashed
line).}
\end{figure}

It is also observed that when the half-bandwidth of the spectrometer $\gamma$ and the coupling constant
between the cavity field and the excitons are of the same order, the number of the peaks in the physical
spectrum (\ref{eq:pppp}) becomes two. This is because the spectrometer can not resolve the position of some
of the peaks. Figure $5$ shows that under the conditions of the weak interaction between the excitons, the
heights of the peaks in spectrum are equal if the cavity field resonates with the excitons. However for the
non-resonant case, the heights of the two peaks are not equal due to the detuning between the frequencies of
the cavity field and that of the excitons even if there is weak interaction between the excitons.

Superposition of  two different exciton states is yet another interesting case to investigate. If we consider
the excitons are initially in the state $\frac{1}{\sqrt{2}}(|1\rangle+|2\rangle)$ and the cavity field is in
the vacuum state $|0\rangle$, then the physical spectrum can be written as
\begin{eqnarray}
S(\omega)&=&\sum_{j=\frac{1}{2}}^{1}\sum_{l,m}\frac{2\gamma}{\gamma^{2}+
(\omega-\omega_{jl,j-\frac{1}{2}m})^{2}} \nonumber \\
&\times&|\langle\psi(0)|\psi_{jl}|^{2}|\langle\psi_{jl}|b^{\dagger} |\psi_{j-\frac{1}{2}m}\rangle|^{2},
\label{eq:two}
\end{eqnarray}
from  which it can be found that there are  eight peaks in the emission spectrum of the excitons. Figure $6$
depicts the line profile of the physical spectrum $S(\omega)$ as a function of the frequency difference
$\omega-\Omega$ under the resonant condition.
\begin{figure}
\epsfxsize=6.5 cm \centerline{\epsffile{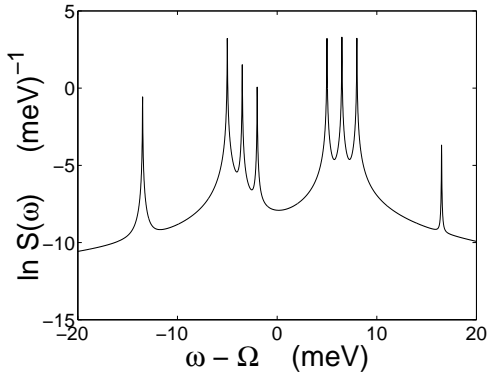}}
\caption[]{$S(\omega)$ is ploted as a function of the\\ 
frequency $\omega-\Omega$ for $g=5$ meV, $A/g=0.6$ \\ 
and $\gamma=0.01$ meV.}
\end{figure}

It is seen from Fig.$6$ that the physical spectrum is a simple sum of one and two exciton spectra. At first
sight,it seems that it is a normal and expected result. However, if the half-bandwidth of the spectrometer
gradually increases,  the peaks of the physical spectrum given by eq.(\ref{eq:two}) reduces from eight to
one.

\section{conclusions}

In this study, the interaction between the high-density excitons
in a  quantum well and the cavity field is investigated under both
the resonant and  non-resonant conditions. The model and the
discussions presented in this paper are valid only when  the
excitons and cavity field  have zero linewidth.

An analytical expression of the physical  spectrum of the excitons is obtained. A discussion of the physical
spectrum of the excitons which are initially prepared in the number state or the superposed state of the two
different number states is presented. It is observed that for a system having a single exciton state
initially,  the resonant interaction between the cavity field and the excitons gives the results similar to
those of the two-level atomic system for which the two peaks in the physical spectrum have equal heights.
However, non-resonant interaction results in the detuning which
 will change both the amplitudes of peaks and the frequency difference
between them.

It is understood that the phase space filling effect does not affect the line profile of the physical
spectrum, but it  adjusts the heights and positions of the physical spectrum.  It is also shown that the
number of the peaks in the physical spectrum is reduced with increase in either the detuning quantity or the
half-bandwidth of the spectrometer. Under the conditions of the low exciton density and the resonant
interaction between the cavity field and the excitons, if the half-bandwidth of the spectrometer has the same
order as the coupling constant between the cavity field and excitons, then spectrum has two peaks with  equal
heights. But in the case of higher exciton density or the non-resonant interaction, even if the the
half-bandwidth of the spectrometer has the same order as the coupling constant between the cavity field and
excitons, the two peaks have different heights due to the interaction between the excitons or the detuning
effect. When the system is initially in the superposed state of a single exciton state and two exctions
state, the resultant physical spectrum  is a simple sum of the spectra of the components of the superposition
states.

\section{acknowledgments}
The authors thank M. Koashi, A. Miranowicz and T. Yamamoto for
helpful and simulating discussions. Yu-xi Liu is supported by
Japan Society for the  Promotion of Science (JSPS). C. P. Sun is
supported by NSF of China. This work also is supported by
Grant-in-Aid for Scientific Research (B) (Grant No.~12440111) by
Japan Society for the Promotion of Science.

\end{multicols}
\end{document}